\documentclass[12pt,preprint]{aastex}
\newcommand{\reff}{\mbox{$R_{\rm e}$}}
\newcommand{\wcc}{\mbox{FWHM$_{\rm cc}$}}

\newcommand{\ishape}{{\tt ishape}}
\newcommand{\tinytim}{{\tt TinyTim}}

\newcommand{\vi}{\mbox{$V\!-\!I$}}

\newcommand{\msun}{\mbox{${\rm M}_{\odot}$}}
\begin{document}
\title{Structure and mass of a young globular cluster in NGC~6946
 \footnote{Based on observations with the NASA/ESA Hubble Space 
 Telescope and with the W. M. Keck Telescope.}
}
\author{S{\o}ren S. Larsen and Jean P. Brodie
  \affil{UC Observatories / Lick Observatory, University of California,
         Santa Cruz, CA 95064, USA}
  \email{soeren@ucolick.org}
\and
  Bruce G. Elmegreen
  \affil{IBM Research Division, T.J. Watson Research Center,
    P.O. Box 218, Yorktown Heights, NY 10598, USA}
\and
  Yuri N. Efremov
  \affil{Sternberg Astronomical Institute, MSU, Moscow 119899, Russia}
\and
  Paul W. Hodge
  \affil{Astronomy Department, University of Washington, Box 351580,
  Seattle, WA 98195--1580, USA}
\and
  Tom Richtler
  \affil{Grupo de Astronom\'{\i}a, Departamento de F\'{\i}sica,
         Casilla 160-C, Universidad de Concepci\'on, Concepci\'on, Chile}
}

\begin{abstract}
  Using the Wide Field Planetary Camera 2 on board the Hubble Space Telescope,
we have imaged a luminous young star cluster in the nearby spiral galaxy 
NGC~6946. Within a radius of 65 pc, the cluster has an absolute visual 
magnitude $M_V=-13.2$, 
comparable to the most luminous young `super-star clusters' in the
Antennae merger galaxy. $UBV$ colors indicate an age of about 15 Myr.
The cluster has a compact 
core (core radius $\sim 1.3$ pc), surrounded by an extended envelope 
with a power-law luminosity profile.  The outer parts of the cluster 
profile gradually merge with the general field, making it difficult to 
measure a precise half-light radius (\reff), but we estimate $\reff\sim13$ pc.
Combined with population synthesis models, the luminosity and age of 
the cluster imply a mass of $8.2\times10^5\,\msun$ for a Salpeter IMF
extending down to 0.1 \msun . If the IMF is log-normal below $0.4\,\msun$
then the mass decreases to $5.5\times10^5\,\msun$.  Depending on model 
assumptions, the central 
density of the cluster is between $5.3\times10^3 \,\msun\,{\rm pc}^{-3}$ and 
$1.7\times10^4 \,\msun\,{\rm pc}^{-3}$, comparable to other high-density
star forming regions.  We also estimate a dynamical mass 
for the cluster, using high-dispersion spectra from the HIRES spectrograph 
on the Keck I telescope.  The HIRES data indicate a velocity dispersion of 
$10.0\pm2.7$ km/s and imply a total cluster mass within 65 pc of 
$1.7\pm0.9\times10^6\,\msun$.  Comparing the dynamical mass with the mass 
estimates based on the photometry and population synthesis models, the
mass-to-light ratio is at least as high as for a Salpeter IMF extending
down to $0.1\,\msun$, although a turn-over in the IMF at $0.4\,\msun$ is 
still possible within the $\sim1\sigma$ errors. The cluster will presumably 
remain bound, evolving into a globular cluster-like object.

\end{abstract}


\section{Introduction}

  Ever since the presence of ultra-luminous young star clusters in certain
external galaxies was first suspected, the true nature of such objects 
has remained somewhat controversial. It took the spatial resolution of
the \emph{Hubble Space Telescope} to definitively prove that the compact
blue objects in starburst dwarfs like NGC~1705 and NGC~1569
\citep{san78,as85} are indeed star clusters and not merely foreground
stars \citep{ocon94}. Subsequently, similar ``super-star clusters''
or ``young massive clusters'' (hereafter YMCs) have been discovered in
other starburst galaxies, notably in mergers like e.g.\ the 
``Antennae'', NGC~7252 and NGC~3256 \citep{whit93,ws95,zepf99}.
From their luminosities and reasonable estimates of the mass-to-light
ratios, YMCs appear to have masses similar to those of the old globular 
clusters observed around virtually all major galaxies, and there is
thus growing anticipation that the study of these young clusters can
provide important information about how their older counterparts formed.

  One remaining challenge is to verify that YMCs contain enough low-mass
stars to remain bound for a significant fraction of a Hubble time.
Deep HST imaging has recently allowed the stellar population of the R136 
cluster in the LMC to be probed down to about 1.35 \msun\ \citep{sir00},
with some evidence for a flattening of the mass function below $\sim 2$ \msun.
Direct observations of low-mass stars in more distant extragalatic star 
clusters are currently far beyond reach.  \citet{bro98} compared 
features in low-resolution spectra of YMCs in the peculiar galaxy NGC~1275 
with population synthesis models and concluded that their data were best 
explained by models with a lack of low-mass stars. However, the integrated 
light of these young objects is generally dominated by A- and B- type stars 
and cool supergiants and conclusions about low-mass stars, based on
integrated spectra and/or photometry are inevitably quite uncertain and 
model-dependent.  A potentially better way to gain insight into the stellar 
mass function 
of unresolved star clusters is to compare dynamical mass estimates with 
the masses predicted by population synthesis models. Should the dynamical 
masses turn out to be much lower than expected, this would indicate that 
the clusters may lack a significant number of low-mass stars.  This, in 
turn, would imply that such objects are \emph{not} similar to old globular 
clusters and will not survive for anything like a Hubble time, since only 
stars with masses below $1 \msun$ have the required long lifetimes.

  Dynamical masses have, so far, been estimated only for a small number of
YMCs in NGC~1569 and NGC~1705 \citep{hf96a,hf96b}, in M82 \citep{sg00}
and the Antennae \citep{meng01}.  \citet{stern98} concluded that the velocity 
dispersion and luminosity of the luminous cluster NGC~1569A are consistent 
with a Salpeter IMF down to 0.1 \msun, while the cluster NGC~1705A may have 
a flatter IMF slope. The luminous cluster 'F' in M82 appears to have a
somewhat lower velocity dispersion than expected from its luminosity, 
favoring a top-heavy IMF \citep{sg01}.  The $\sim100$ Myr old cluster 
NGC~1866 in the Large Magellanic Cloud was studied by \citet{fis92}, who 
obtained a dynamical mass of $1.35 \times 10^5$ \msun . \citet{van99} pointed 
out that, for the luminosity and age of NGC~1866, this mass implies a very 
high mass-to-light radio and large numbers of low-mass stars.  

  In a study of 21 nearby spiral galaxies, \citet{lr99} found several
examples of YMCs. Most of these galaxies are at distances less than about 
10 Mpc and thus offer attractive targets for detailed studies of their YMC 
populations. A particularly interesting, very luminous young cluster was 
found within a peculiar bubble-shaped star forming region in the nearby 
face-on spiral NGC~6946.  \citet{tul88} lists a distance of 5.5 Mpc and 
more recently a mean distance of $5.9\pm0.4$ Mpc has been estimated for 
the NGC~6946 group from ground-based photometry of the brightest blue stars 
\citep{ksh00}. For the remainder of this paper we adopt the latter distance 
estimate.  The star forming region was first noted by \citet{hod67} in a 
search for objects similar to Constellation III in the LMC, but was then 
largely forgotten.  Using ground-based CCD images from the Nordic Optical 
Telescope, the young massive cluster and its surroundings were further 
discussed by \citet{eel00}, who estimated a total mass of about 
$5\times10^5\,\msun$ and an age of 15 Myr based on $UBV$ colors.  The 
ground-based data, 
obtained in a seeing of about $0\farcs7$, also provided an estimate of the 
half-light (effective) radius of the cluster of about 11 pc. With an 
apparent $V$ magnitude of about 17, it may seem surprising that such a 
luminous object in a nearby galaxy went relatively unnoticed until recently. 
This may be due to the fact that NGC~6946 is located at a low galactic 
latitude ($b = 12\deg$), in a field rich in foreground stars. Therefore, on 
ground-based images taken in less than optimal seeing, the cluster is 
easily confused with a foreground star.

\begin{figure}
\plotone{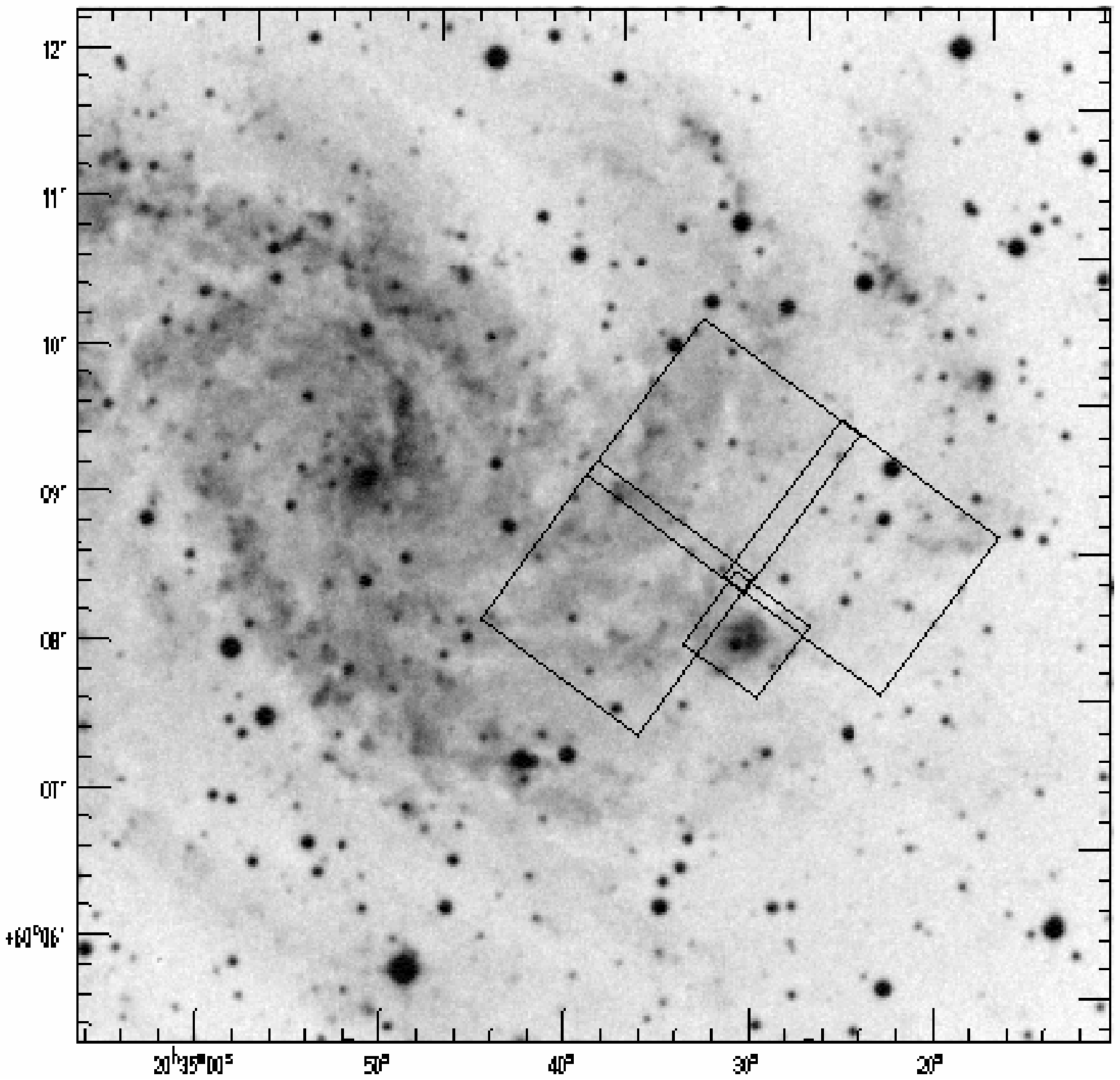}
\figcaption[f1.eps]{\label{fig:fov}The WFPC2 field of view superimposed on a
Digital Sky Survey image of NGC~6946. North is up and East to the 
left in this figure.
}
\end{figure}

  Here we present new HST / WFPC2 and Keck I / HIRES data for the young 
massive cluster in NGC~6946, labeled n6946-1447 by \citet{lar99}.
The WFPC2 field of view is shown in Fig.~\ref{fig:fov}, superimposed on
an image of NGC~6946 from the Digital Sky Survey. A color image from the
Nordic Optical Telescope, showing the region around the cluster, can be
found in \citet{eel00}.  

\section{Data}

\subsection{HST data}

  WFPC2 data were acquired in Cycle 9, using the F336W ($U$), 
F439W $(B)$, F555W $(V)$ and F814W $(I)$ filters. The integration times
were 3000 s, 2200 s, 600 s and 1400 s in the four bands, respectively,
with all integrations split into two exposures in order to facilitate
efficient elimination of cosmic ray hits.  Initial processing (bias
subtraction, flatfielding etc.) were performed ``on-the-fly'' by the 
standard pipeline processing system at STScI. The individual exposures were 
then combined using the IMCOMBINE task within IRAF\footnote{IRAF is 
distributed by the National Optical Astronomical Observatories, which are 
operated by the Association of Universities for Research in Astronomy, 
Inc.\ under contract with the National Science Foundation}, setting the 
{\bf reject} option to {\bf crreject} in order to eliminate cosmic ray 
(CR) hits.

  The PC field of view of the F555W exposure is shown in Fig.~\ref{fig:pc}. 
The whole star forming complex comfortably fits on the PC chip and we do 
not consider data in the Wide Field chips in this paper.  The YMC is
easily recognizable as the single most luminous object in the field.
At the distance of NGC~6946, one PC pixel corresponds to a linear scale
of 1.3 pc so unless the YMC is unusually compact, its radial profile will
be well resolved. In this paper we only discuss the cluster itself and
its immediate neighborhood.  The numerous other clusters and individual 
stars contained within the PC field will be discussed in a subsequent 
paper.

\subsection{Keck data}
 
  Observations with the HIRES high-dispersion spectrograph \citep{vogt94}
on the Keck I telescope were obtained during two half nights in August
2000. Both nights were photometric with a seeing of around $0\farcs9$.
Following \citet{hf96a}, we used two different setups: One optical
setting covering the wavelength range from 3780 \AA\ to 6180 \AA\ in 
37 echelle orders, and a near-infrared setting ranging from
6220 \AA\ to 8550 \AA\ in 16 orders. For the optical setting we used
the ``C1'' decker, providing a $7\arcsec \times 0\farcs86$ slit,
while the ``D1'' decker with a $14\arcsec \times 1\farcs15$ slit
was used for the near-IR setting. This provided a spectral resolution
of $R=45000$ and $R=34000$ for the two settings, respectively.

  The total integration times were 180 minutes and 220 minutes for the
optical and near-IR spectra, split into 4 individual exposures for each
setting.  The slit orientation was kept constant with respect to the sky
during the exposures, but was aligned with the parallactic angle at the
beginning of each exposure. In addition to the cluster
spectra we also obtained spectra for 11 stars of different
spectral types, to be used as cross-correlation templates. These are
listed in Table~\ref{tab:tmpl}.

\begin{figure}
\plottwo{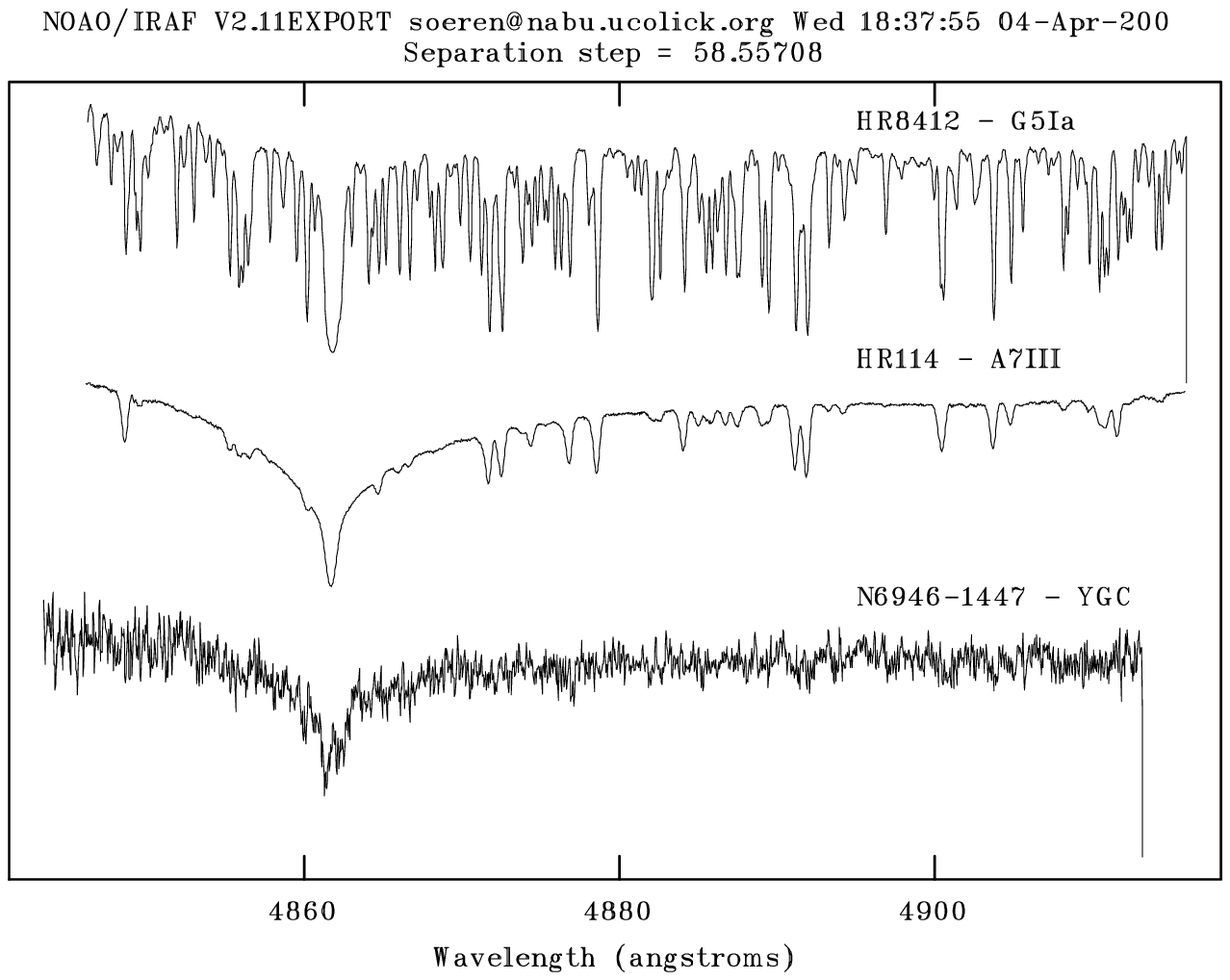}{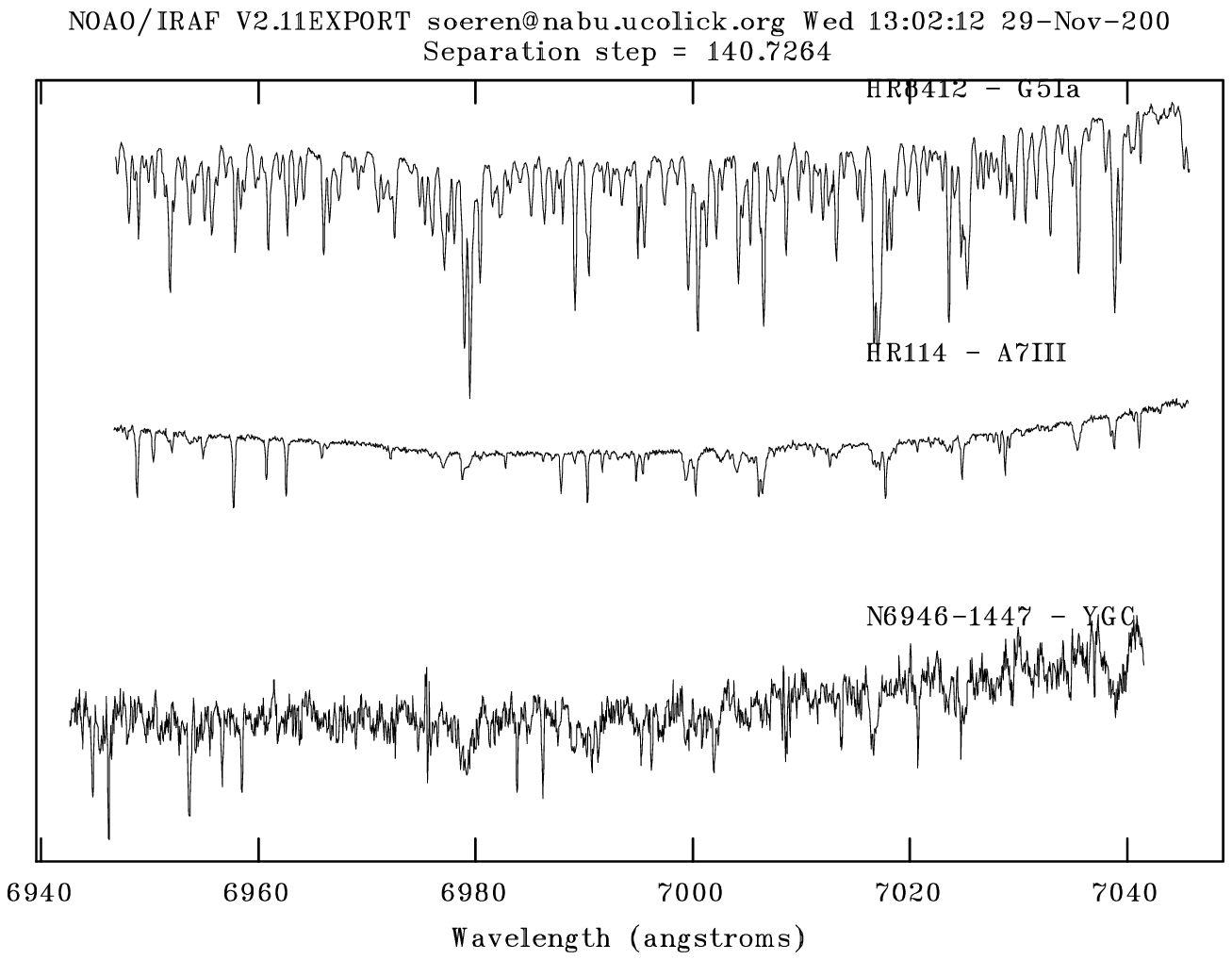}
\figcaption[f3a.eps,f3b.eps]{\label{fig:spec}
Two echelle orders from the HIRES spectra of the young globular cluster
(bottom) and two comparison stars. The spectra have been shifted to
the same wavelength scale, correcting for the radial velocity of
NGC~6946.}
\end{figure}

  The reductions were performed using the highly automated ``makee'' 
package, written by Tom Barlow. ``makee'' automatically performs bias 
subtraction and flatfielding, identifies the location of the echelle orders 
on the images and extracts the spectra. Wavelength calibration is done using
spectra of ThAr calibration lamps mounted in HIRES.  Each of the individual 
1-dimensional
spectra were then combined using the {\bf scomb} task in IRAF.  In 
Fig.~\ref{fig:spec} we show two echelle orders from the cluster spectra
compared with a G5Ia star (HR 8412) and an A7III star (HR 114). The left
plot includes the $H\beta$ line.  The cluster spectrum is clearly of a 
composite nature, showing both strong 
Balmer lines similar to those in early-type stars, and numerous lines 
due to heavier elements, as in evolved cool supergiants. 

\section{Results}

\subsection{Structure of the cluster}
\label{sec:struct}

\begin{figure}
% \epsfxsize=10cm
% \epsfbox{f2sm.ps}
\plotone{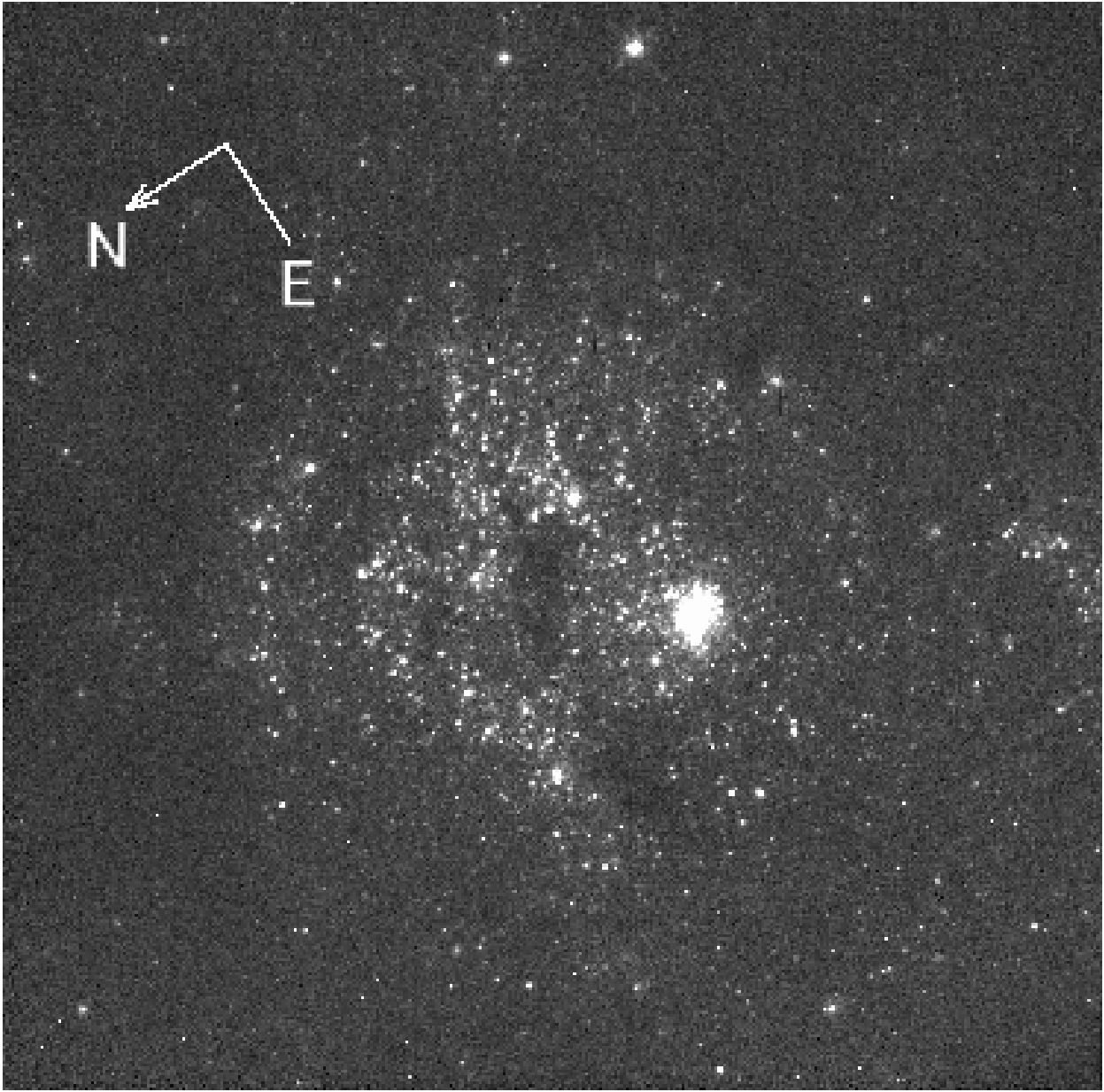}
\figcaption[f2.eps]{\label{fig:pc}An F555W image showing the PC field of view.
North and East are indicated by the arrow.  The young globular cluster 
is easily recognizable as the single most luminous object, near the
center of the image.}
\end{figure}

\begin{figure}
\plotone{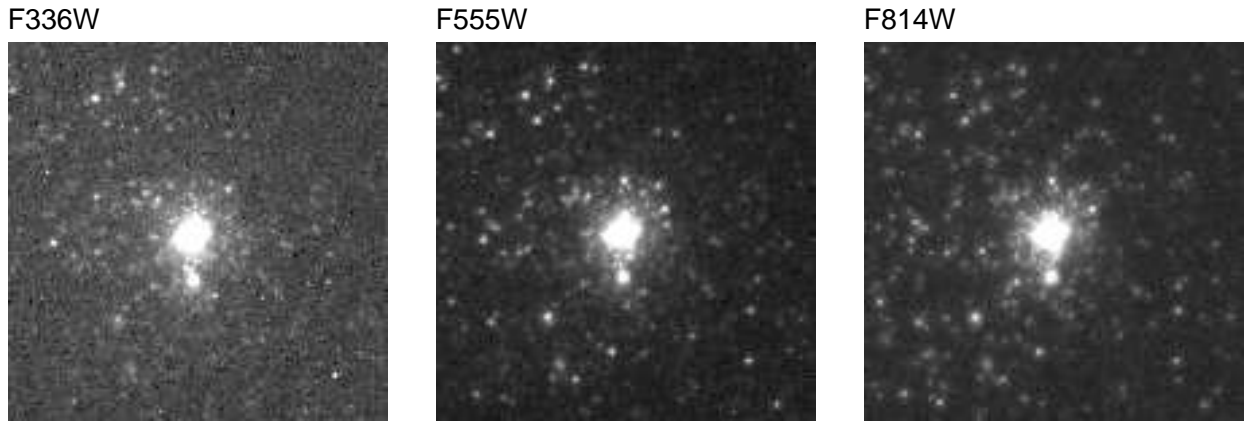}
\figcaption[f4.eps]{\label{fig:ygc}Close-ups of the young globular in
F336W, F555W and F814W. Each image spans about 6\arcsec . Orientation
is the same as in Fig.~\ref{fig:pc}.
}
\end{figure}

  Fig.~\ref{fig:ygc} shows close-ups of the cluster in F336W, F555W
and F814W, each spanning 6\arcsec\ or about 170 pc across. Note that
another fainter cluster is located about 15 pixels (19 pc) to the
North-East. In addition a number of field stars are visible, in
particular in the F814W image, and also the young globular itself begins 
to resolve into individual stars. From Figs.~\ref{fig:pc} and \ref{fig:ygc}, 
it is hard to tell where the cluster ends and where the general field 
population begins. In fact, star formation appears to have taken place 
over an area much larger than that directly connected with the cluster, and 
only the stars closest to the cluster center may actually be bound to it.
% !!! changed to "closest to the cluster center" above because stars on the periphery
% of the cluster may not be bound to it. 

\begin{figure}
\plottwo{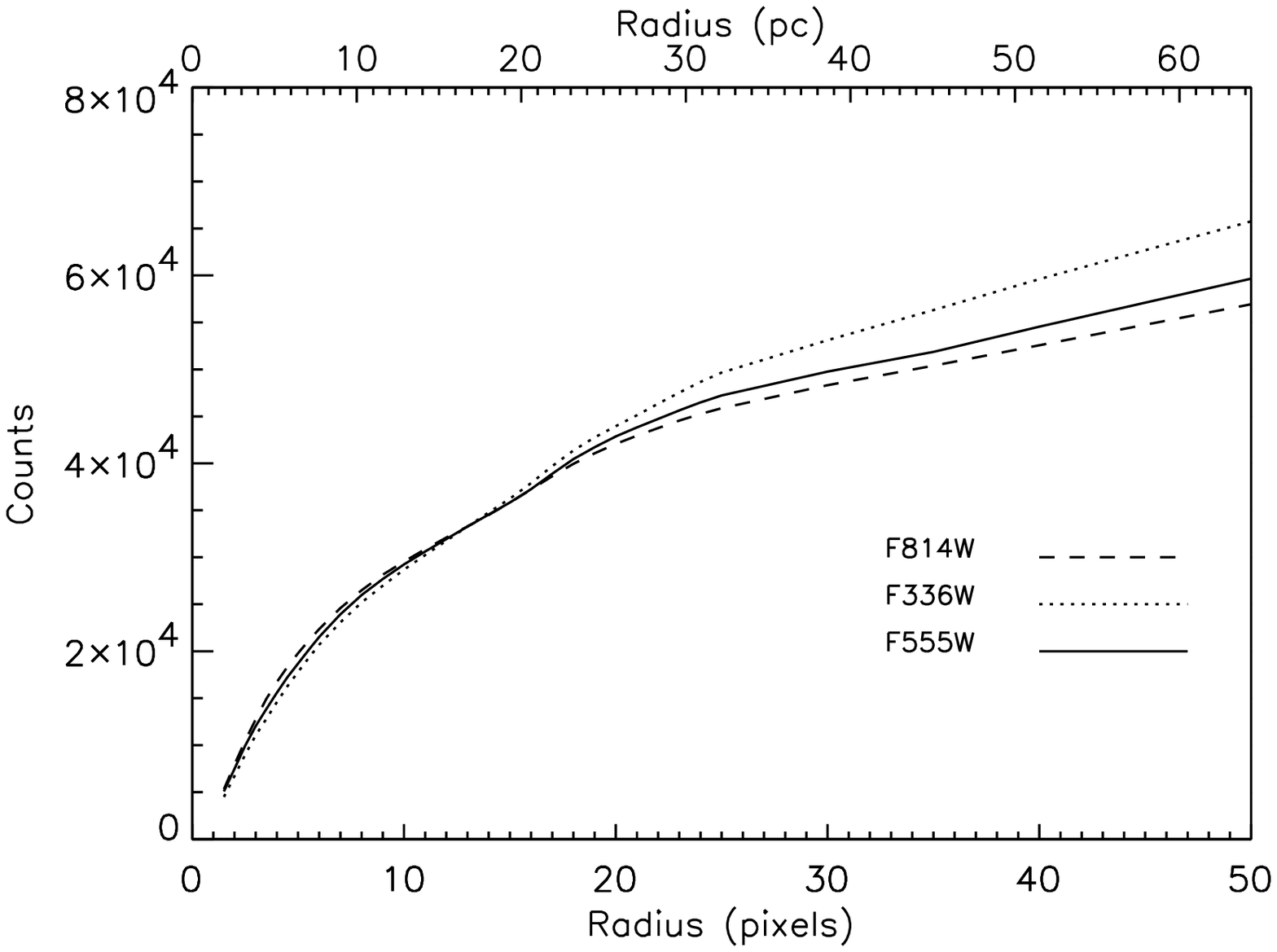}{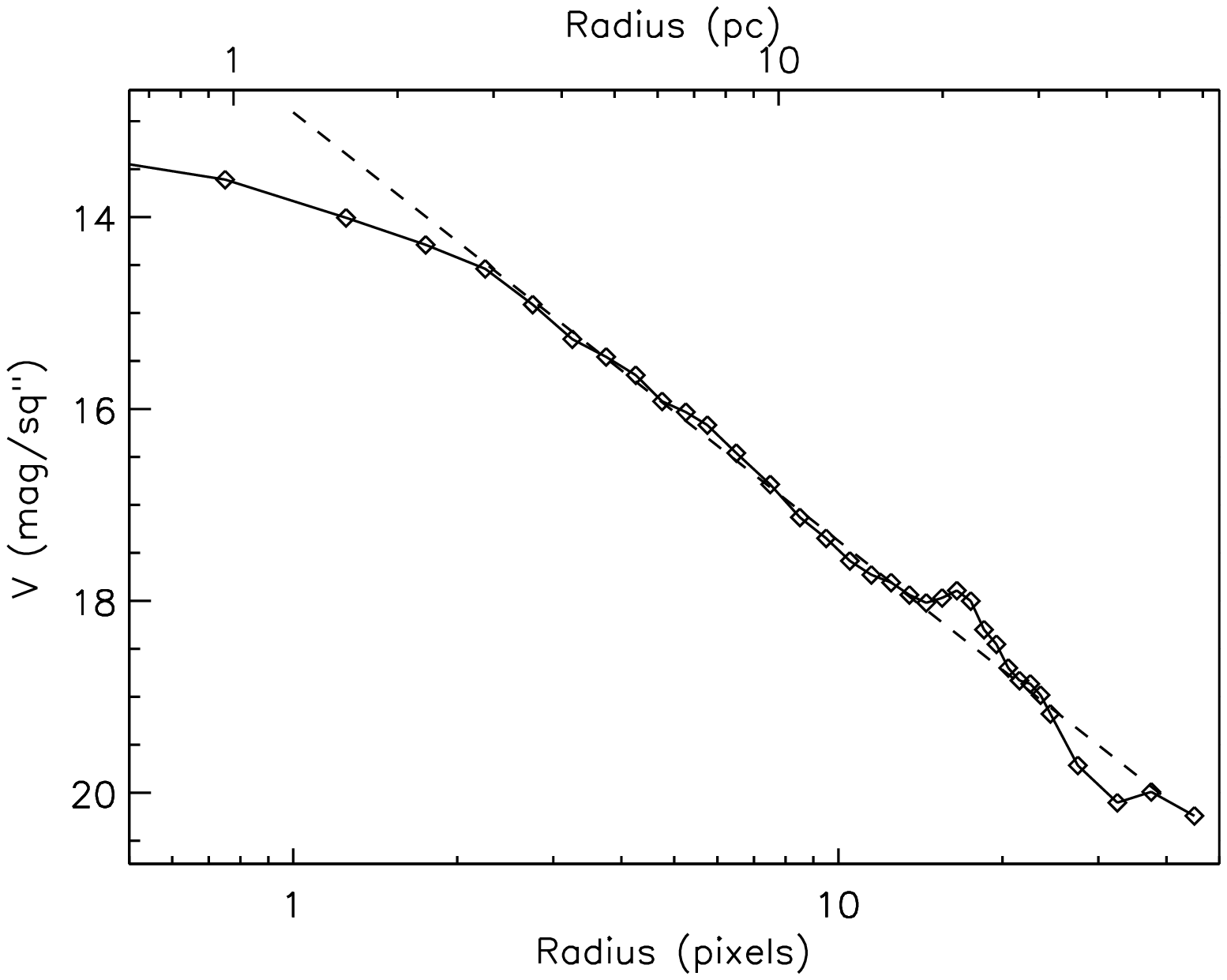}
\figcaption[f5a.eps,f5b.eps]{\label{fig:rprof}
  Left: Luminosity of the cluster as a function of 
radius, measured on F336W, F555W and F814W images. The F336W and F814W 
profiles have been normalized to that measured on the F555W image. 
Right: The $V$-band surface brightness profile, corrected for foreground
extinction.  Note the `bump' in the
profile at $r\sim15$ pixels, resulting from the fainter companion
cluster. The power-law envelope clearly extends to very large radii.
One pixel equals $0\farcs045$ or about 1.3 pc. The dashed line is a
power-law fit.  No correction for the instrumental profile has been made 
in this figure.
}
\end{figure}

  In Fig.~\ref{fig:rprof} we show the integrated luminosity (left) and
surface brightness (right) of the cluster as a function of 
radius. The photometry was done using the PHOT task within the DAOPHOT
package in IRAF, measuring the background in an annulus starting at 50 pixels 
($2\farcs25$) and 50 pixels wide.  The instrumental profile has not been taken 
into account in Fig.~\ref{fig:rprof}.  We do not measure the luminosity
profile of the cluster beyond 50 pixels, as it is clear e.g.\ from 
Fig.~\ref{fig:pc} that the irregular background at larger distances from 
the cluster would make such a measurement very uncertain. Nevertheless,
the integrated light will probably continue to rise well beyond 50 
pixels ($\sim 65$ pc).  This is not unusual for young 
star clusters: \citet{els87} found that young (8 -- 300 Myr) clusters in the 
LMC are surrounded by large envelopes with power-law luminosity profiles
which will probably be lost to tidal forces. Similarly, \citet{whi99} found 
an extended envelope with a diameter of more than 900 pc around the highly 
luminous ``knot S'' in the Antennae. 

  If we approximate the surface brightness $(SB)$ as a function of radius 
$r$ with a simple power-law of the form $SB \propto r^{\alpha}$ 
(for $SB$ in counts / unit area) and perform a fit to the cluster profile 
between $r=2$ and $r=15$ pixels, we formally obtain an exponent of 
$\alpha = -1.79\pm0.03$. The fit is indicated by the dashed line in the 
right-hand panel of Fig.~\ref{fig:rprof}. 

  Although a power-law may provide a satisfactory fit to the outer parts
of a cluster, the intensity must level off at some radius near the
center. Thus, a more realistic analytic model of the cluster profile will
involve some core radius $r_c$.  \citet{els87} found that the surface
brightness profiles of young clusters in the LMC are generally well fit 
by analytic models of the form
\begin{equation}
  SB \, \propto \, (1 + r^2/r_c^2)^{-\xi/2}
  \label{eq:els87}
\end{equation}
with exponent $\xi$ in the range $2.2 < \xi < 3.2$ and core radii 
from 1.3 to 7 pc. For $r\gg r_c$, equation~(\ref{eq:els87}) is similar
to a power-law with slope $-\xi$, but reaches a constant value near the 
center.

  Because the core radius of n6946-1447 is comparable to the
resolution of the PC camera, it cannot be accurately estimated from
a simple plot of surface brightness vs.\ radius.  We have used a modified 
version of the \ishape\ algorithm \citep{lar99} to fit models of the form 
(\ref{eq:els87}) to the image of the young globular. \ishape\ iteratively 
adjusts the exponent $\xi$ and the core radius $r_c$ and then convolves 
the corresponding model with the HST point-spread function (PSF) until the 
best match to the observed cluster image is obtained. The HST PSF is 
modeled using the \tinytim\ PSF simulator \citep{kri97} and the modeling
done by \ishape\ also involves a convolution with the WFPC2
``diffusion kernel'' \citep{kri97}. Since the diffusion kernel is best
characterized for the F555W band, we used exposures in this band for the
model fits.

  To test the stability of the fitted parameters we carried out
a number of fits, varying the fitting radius between 5 and 15 pixels
and changing the initial guesses for the exponent $\xi$. The algorithm 
returned exponents in the range $1.98<\xi<2.18$ and FWHM values between
1.70 and 2.19 pixels. We thus adopt FWHM = $1.95\pm0.25$ pixels and 
$\xi = 2.1\pm0.1$ as our estimates of the structural parameters for 
the cluster. Note that this is a slightly steeper profile than that
obtained by a simple power-law fit to the raw cluster profile, uncorrected 
for instrumental effects.  For the relevant range of $\xi$ values, the core
radius $r_c = 0.5\times\mbox{FWHM}$ to within 5\%, i.e.\ $r_c = 1.26\pm0.16$ pc 
for a distance of 5.9 Mpc.
This is comparable to the most compact young LMC clusters and is also a 
typical value for Milky Way globular clusters \citep{pk75,har96}. In 5 of the 
6 fits we performed, \ishape\ returned minor/major axis ratios between 
0.91 and 0.93, while one fit (for $r=5$ pixels) returned an axis ratio of 
0.97. The cluster thus seems to be somewhat elongated with an axis ratio 
of about 0.92, with a likely uncertainty of a few times 0.01.

  Integrating equation~(\ref{eq:els87}) from $r=0$ to $r=\infty$, the total 
luminosity diverges for $\xi<2$. With our estimate of 
$\xi = 2.1\pm0.1$, the half-light radius (\reff ) is therefore not 
very well-defined.  For $\xi = 2$, (\ref{eq:els87}) is identical 
to a \citet{king62} profile with infinite tidal radius. We also
attempted to fit King models to the cluster profile, varying the 
concentration parameter, but such fits are highly sensitive to 
inaccuracies in the background level and turned out to be too 
uncertain.  If we (somewhat arbitrarily) define the ``total'' luminosity 
as the luminosity contained within 50 pixels, Fig.~\ref{fig:rprof} suggests 
$\reff \sim 10$ pixels or about 13 pc.  This crude estimate is 
significantly larger than the typical half-light radius for stellar 
clusters, but agrees well with the ground-based estimate of 
$\reff \sim 11$ pc obtained by \citet{eel00}.  Old globular clusters 
typically have $\reff\sim3$ pc \citep[e.g.][]{har96}, while young massive
clusters in starburst / merger galaxies such as the Antennae may have
slightly larger effective radii \citep[$\reff\sim4$ pc, ][]{whi99}.  
However, because of the youth of the cluster in NGC~6946, it 
is quite likely that much of the loosely bound outer parts may eventually 
be stripped.  Assuming that the core remains relatively unaffected by tidal 
stripping, we can calculate the half-light radius for King models 
with various tidal radii ($r_t$).  Integration of the King profiles shows 
that the effective radius \reff\ and core radius are related as 
$\reff =  2.9 \, r_c$ for $r_t/r_c = 30$ and $\reff = 5.1 \, r_c$ for 
$r_t/r_c = 100$.  Assuming $r_c = 1.3$ pc for n6946-1447, we then obtain 
$\reff = 3.8$ pc and $\reff = 6.6$ pc for $r_t/r_c = 30$ and 
$r_t/r_c=100$, respectively. These numbers suggest that, if the cluster
evolves towards a King profile with a finite tidal radius, its effective
radius could decrease significantly.

  It is also worth noting that some old globular clusters have significantly 
larger half-light radii than 3 pc. The \citet{har96} catalog lists 
$\reff$ values up to about 20 pc for some of the outer Palomar-type halo 
clusters, and \citet{hph98} obtained a half-mass radius of about 7 pc for 
a globular cluster in NGC~5128, corresponding to $\reff \sim 5.3$ pc.

\subsection{Integrated photometry}
\label{sec:phot}

  In Table~\ref{tab:phot} we list photometry for the young cluster in 
5 different apertures between $r=5$ pixels ($0\farcs23$) and $r=50$ pixels
($2\farcs25$).  Again, the photometry was obtained using the PHOT task 
within IRAF, measuring the background between 50 and 100 pixels from the 
cluster center.  Instrumental magnitudes measured on the PC images were 
transformed to the standard $UBVI$ system using the transformations in 
\citet{hol95}.  For comparison, we also list
the ground-based photometry from \citet{lar99}. The magnitudes and
colors in Table~\ref{tab:phot} have not been corrected for galactic
foreground extinction, and no aperture corrections have been applied to
the HST data other than the $-0.1$ mag correction which is implicit in
the Holtzman et al.\ calibration. Strictly speaking, the Holtzman et al.\
calibration is only valid for a point source observed through a 
$0\farcs5$ aperture (11 pixels). In order to get the ``true'' magnitude
for the cluster, we would have to observe it through an
aperture known to encompass 90\% of the total luminosity. Comparing
with Fig.~\ref{fig:rprof}, it is quite likely that not even our
$r=50$ pixels aperture contains 90\% of the cluster light, so the
true total magnitude of the cluster may be even brighter than $V=16.7$.

  Both Fig.~\ref{fig:rprof} and Table~\ref{tab:phot} show that
the colors are somewhat bluer when measured through the larger apertures. 
To test if this could be due to wavelength-dependent aperture corrections,
we convolved \tinytim\ PSFs in different bands with a cluster model of 
the form (\ref{eq:els87}) 
and carried out photometry in the same apertures as those listed in 
Table~\ref{tab:phot}.  Two sets of tests were performed, one with the sky 
background kept at a fixed level of 0 and another where the background was 
measured on the synthetic images in the same annulus as for the real 
photometry.  For \ub\ and \bv, 
we found no change larger than $\sim0.01$ mag in the color index aperture 
corrections from 5 to 50 pixels, while the tests show that the \vi\ colors 
are about 0.03 mag too blue when measured through the $r=5$ pixels aperture.  
These results are essentially independent of how the background was measured.
The color gradient thus appears to be real, although it is not clear 
whether it is intrinsic to the cluster or due to differential reddening,
contamination by field stars with different ages, or other causes.

  Correcting the $r=50$ pixels photometry for a Galactic foreground 
extinction of $A_B=1.48$ \citep{sch98} and using the reddening law by 
\citet{car89}, we get $(\ub)_0 = -0.75$ and $(\bv)_0 = 0.04$.  Adopting 
the \citet{gir95} `S-sequence' calibration for age as a function of 
$UBV$ colors, this implies a cluster age of about 15 Myr, identical to
the value reported by \citet{eel00} based on ground-based photometry. 
Our 15 Myr age estimate is also compatible with a spectroscopic age
determination, based on Balmer line equivalent widths measured on
low-dispersion spectra from the 6 m Special Astrophysical Observatory 
(SAO) telescope (Efremov et al., in
preparation). Furthermore, the SAO data (as well as the Keck/HIRES
spectra) show no H$\alpha$ emission from the cluster itself, indicating a 
lower bound on the age of $\sim 10$ Myr.  \citet{gir95} quote an rms
scatter of 0.137 in log(age) for the S-sequence calibration, corresponding
to an uncertainty of about $\pm5$ Myr for a 15 Myr old cluster.  We note 
that, since the S-sequence age calibration is based on LMC clusters, it 
may not be strictly valid for clusters with different metallicity.  However, 
$UBV$ colors are not very sensitive to metallicity below 
$\sim 500\times10^6$ yr so this should not lead to any large errors in 
the age estimate. 

\subsection{Velocity dispersion}

  From the virial theorem, the total mass ($M$), velocity dispersion
($v$) and half-mass radius ($r_h$) of an isolated cluster with
isotropic velocity distribution are related as
\begin{equation}
  M = a \frac{v^2 r_h}{G},
  \label{eq:virial}
\end{equation}
where the constant $a$ has a value of about 2.5 \citep[p.\ 11]{spi87}. Note 
that $v$ is the 3-dimensional velocity dispersion. What we actually
measure is the line-of-sight velocity dispersion, $v_x^2 = \frac{1}{3} v^2$. 

  The spectral resolution $R\sim40000$ of the HIRES spectra is of the 
same order of magnitude as the expected velocity dispersion within the 
cluster, so simply measuring the width of the spectral lines directly on 
the spectra would not provide a realistic estimate of the true velocity 
dispersion. Instead, we follow the same procedure used by 
\citet{hf96a,hf96b}. First, the cluster spectrum is cross-correlated with 
the spectra of a number of suitable template stars, using the {\bf fxcor} 
task in IRAF. The velocity dispersion $v_x$ is then obtained from the 
full width at half maximum of 
the cross-correlation peaks, \wcc . The relation between \wcc\ and $v_x$
is established empirically by artificially broadening the template star 
spectra by convolution with Gaussian profiles and then cross-correlating the
broadened template spectra with the spectra of other template stars.
\wcc\ is independent of image noise,
although the height of the peak does depend on the signal/noise of
the cross-correlated spectra. For very poor S/N spectra the cross-correlation
peak vanishes into the noise. Our cluster spectra typically have a
S/N of about 50, but by artificially adding noise to our spectra we found 
that consistent measurements of \wcc\ were still possible even if the 
S/N of the cluster spectra was degraded to below 10.

\begin{figure}
\plotone{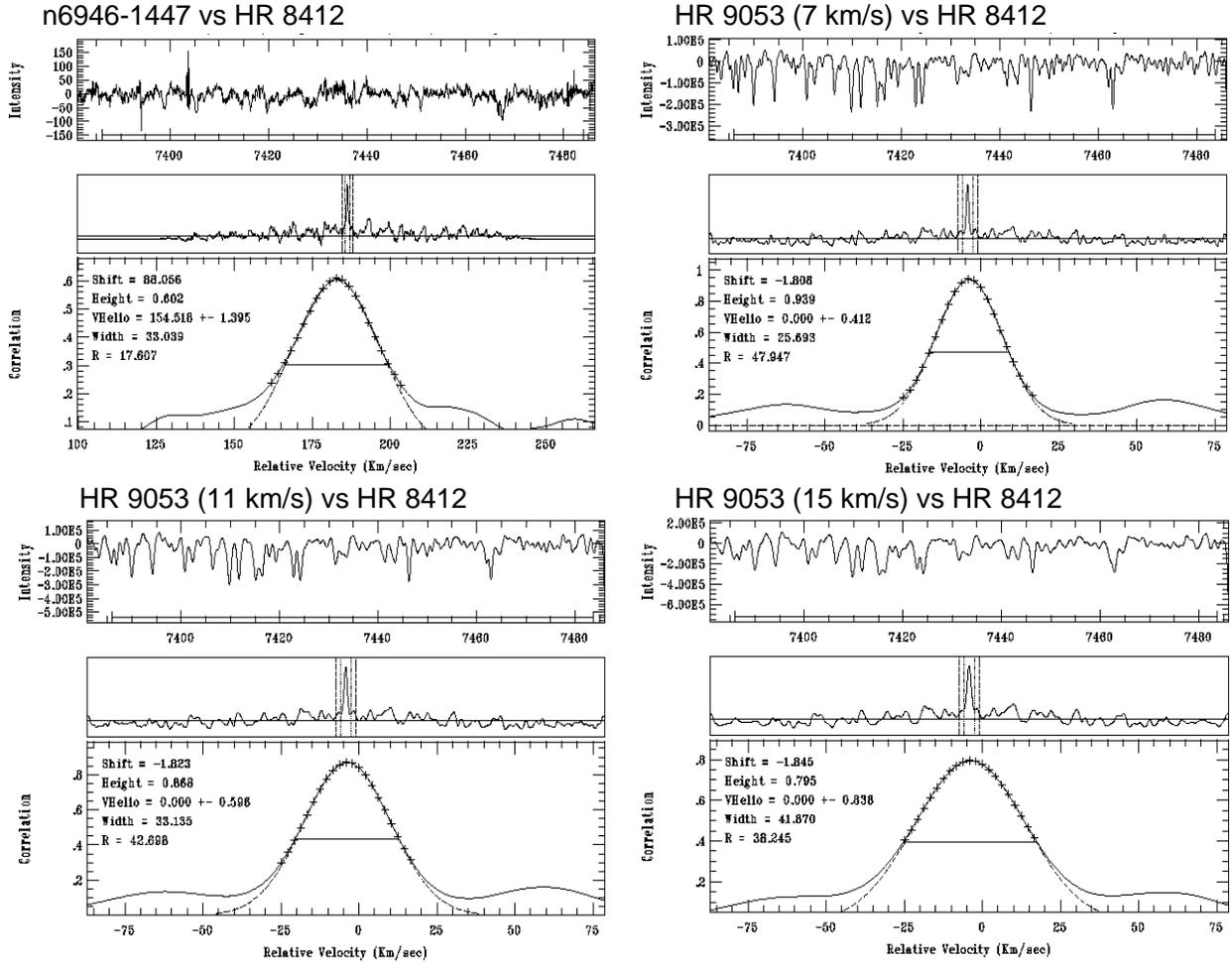}
\figcaption[f6.eps]{\label{fig:ccorr}Cross-correlation functions for
the young cluster vs.\ HR 8412 (upper left) and HR 9053 vs.\ HR 8412
where the HR 9053 spectrum has been convolved with Gaussians
corresponding to velocity dispersions of 7 km/s, 11 km/s and 15 km/s.
}
\end{figure}

  The principle is illustrated in Fig.~\ref{fig:ccorr}: The upper
left-hand panel shows the cross-correlation peak for one echelle order
of the cluster spectrum vs.\ the template star HR 8412. The remaining
panels show cross-correlation peaks for the template star HR 9053,
convolved with Gaussians corresponding to $v_x=7$ km/s, $v_x=11$ km/s
and $v_x=15$ km/s vs.\ the same template star as for the cluster spectrum.
For the cluster vs.\ HR 8412 peak, \wcc\ = 33 pixels, compared
to \wcc\ = 25.7, 33.1 and 41.9 pixels for the three test cases.
In this case, the data imply a velocity dispersion for the cluster
spectrum close to 11 km/s. 

\begin{figure}
\plotone{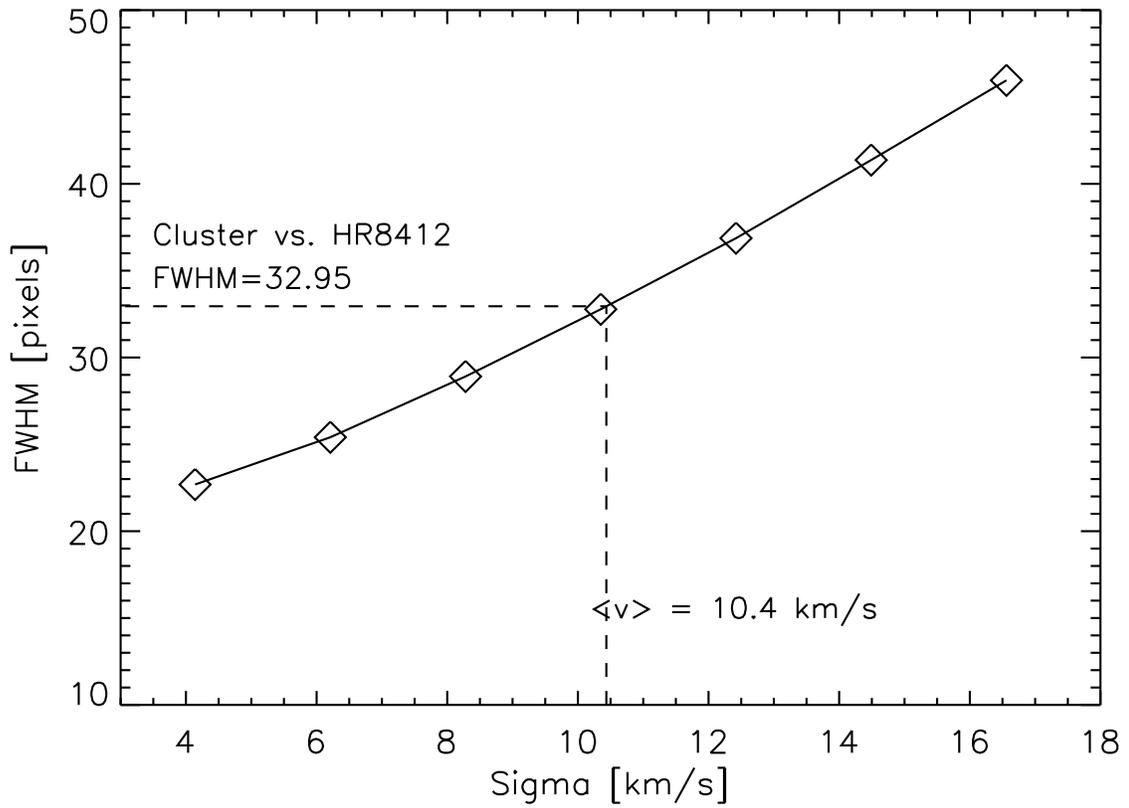}
\figcaption[f7.eps]{\label{fig:demo}
  Illustration of the technique used for deriving
 velocity dispersions: The diamonds indicate the FWHM of the cross-correlation
 peak for the template star HR 8412 vs.\ HR 9053, the latter convolved with
 Gaussians of different dispersions as indicated on the $x$-axis of the
 plot. The FWHM of the cluster vs.\ HV 8412 cross-correlation peak
 of 32.95 pixels corresponds to a velocity dispersion of 10.4 km/s.  
}
\end{figure}

  In practice, we convolved the template star spectra with a number
of Gaussians, having dispersions between 2 and 8 pixels (4.15 km/s
to 16.6 km/s). In this way, \wcc\ was empirically established 
as a function of $v_x$ for each combination of template stars. In 
Fig.~\ref{fig:demo}
this is illustrated for just one combination of template stars, 
HR 8412 and HR 9053 (as in Fig.~\ref{fig:ccorr}). The diamonds show 
\wcc\ for the HV 9053 spectrum convolved with Gaussians of different 
velocity dispersions. Measurements where \wcc\ for the cluster 
vs.\ template star peak fell outside the range corresponding to a velocity 
dispersion in the 4.15 -- 16.6 km/s interval were rejected.

  As is evident from Fig.~\ref{fig:spec}, the saturated nature of the 
Balmer lines, along with the rapid rotation of hot early-type stars, makes 
them unsuitable for
measurement of velocity dispersion. We therefore used regions of the
spectra dominated by features from cool supergiants.  Cool supergiants
in a $\sim15$ Myr old cluster are expected to be of luminosity class 
Ia -- Ib, but to test the sensitivity of the results to the luminosity 
class of the
templates, we observed a number of template stars covering a range
of spectral types as well as luminosity classes. The template stars
are listed in Table~\ref{tab:tmpl}.

\begin{figure}
\plotone{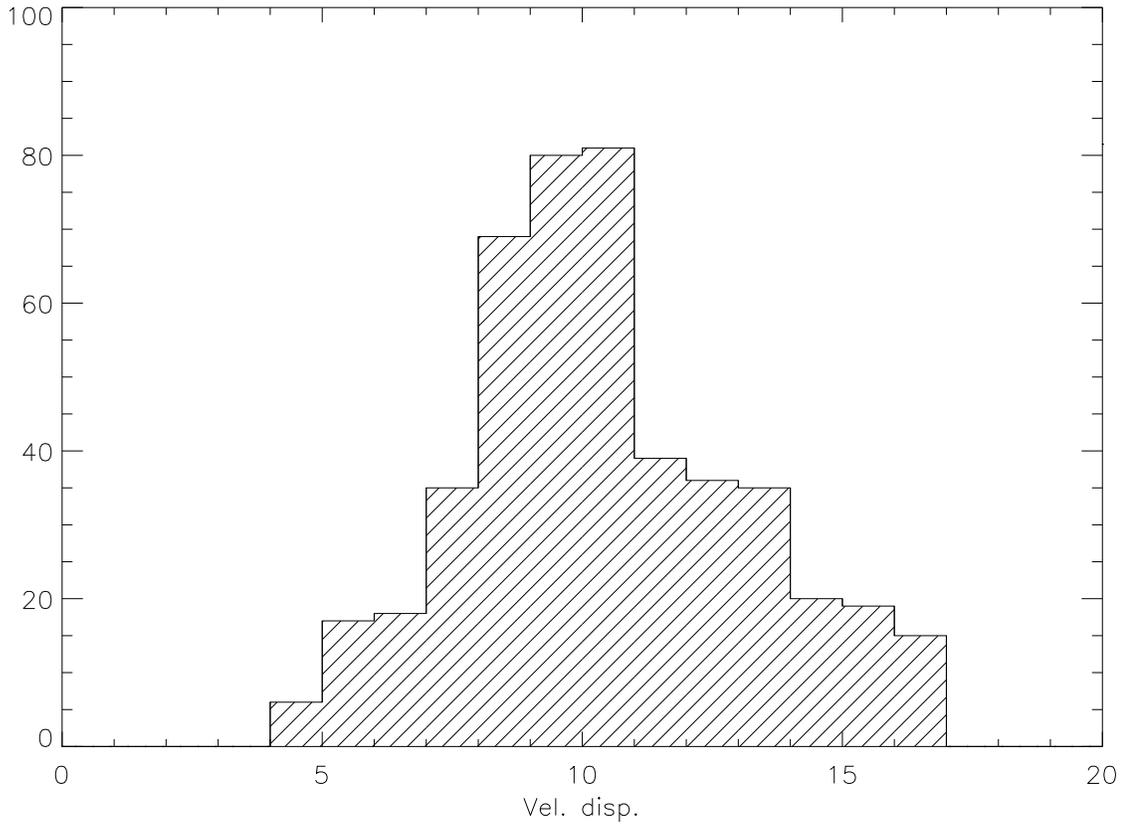}
\figcaption[f8.eps]{\label{fig:vdisp}The distribution of velocity dispersions
derived from all combinations of template stars and echelle orders.
}
\end{figure}

  In spite of the large number of echelle orders, only relatively few turned 
out to be useful. In the optical setting, signal-to-noise was quite low at 
the blue end and only 3 orders were used. In the near-infrared setting, many 
orders were dominated by skylines, reducing the number of useful orders to 6.
Even so, the number of 
individual ``measurements'' of the cluster velocity dispersion is very 
large, each ``measurement'' consisting of cross-correlation of a cluster 
spectrum with a template star and comparing with the template star 
cross-correlated with another smoothed template star spectrum. The 
distribution of all velocity dispersion measurements is shown in 
Fig.~\ref{fig:vdisp}, and Table~\ref{tab:vd} lists the median velocity 
dispersion for each echelle order for stars of luminosity classes I, II 
and III separately.  In some cases, only part of an order was used, as 
indicated by column 2 of Table~\ref{tab:vd}.
As can be seen from Fig.~\ref{fig:vdisp}, the lower
and upper velocity dispersion limits bracket the relevant range quite
well. The median value is 10.1 km/s and the standard deviation 
is 2.7 km/s. 

  The velocity dispersions are generally larger when using template stars 
of luminosity class III.  This is not surprising, considering the significant 
amounts of 
macro-turbulence in the atmospheres of cool supergiants which contributes to 
line broadening \citep{gt87}. Thus, when using normal cool giants as 
templates, the velocity dispersion of the cluster stars will be overestimated.

  Using only template stars of luminosity class I, the median value for the
velocity decreases slightly to 10.0 km/s, while the scatter remains
at 2.7 km/s.  As a further check, we also computed the velocity dispersion
using the three best-fitting templates. As determined from the height
of the cross-correlation peaks, these are HR 1009 (M0II), HR 861 (K3Ib)
and HR 8726 (K5Ib). The best fits were obtained from orders 10 and 14 in
the IR setting, leading to an average velocity dispersion of 9.4 km/s with a
standard deviation of 0.57 km/s for this best-fitting template subsample.
This value is slightly lower than that based on the full
sample, but within the error margins there is good agreement.  We thus 
adopt $v_x = 10.0 \pm 2.7$ km/s as our final estimate of the velocity 
dispersion of n6946-1447, noting that the uncertainty estimate is 
probably quite conservative.

\subsection{Cluster mass}

\subsubsection{Dynamical mass}
\label{sec:mdyn}

  With an estimate of the cluster velocity dispersion and physical size
at hand, we are now ready to estimate the dynamical mass. Note that the 
cluster radius used in equation~(\ref{eq:virial}) is the 3-dimensional 
\emph{half-mass} radius $r_h$, which is larger than the 2-dimensional 
\emph{half-light} or effective radius $\reff$ measured on the images by 
approximately a factor of 1.3 \citep{spi87}. As discussed in 
Sect.~\ref{sec:struct}, the effective radius of the cluster is not very 
well determined, but if we tentatively adopt $r_h = 17$ pc and insert 
this number, together with a velocity dispersion of $v_x = 10.0\pm2.7$ 
km/s in equation~(\ref{eq:virial}), then the total virial cluster mass 
becomes $3.0\pm1.6\times10^6 \, \msun$. 

  However, obtaining the dynamical cluster mass directly from 
equation~(\ref{eq:virial}) is inaccurate for a number of reasons: As already
mentioned, the half-mass radius is uncertain because we only observe
the cluster profile out to 50 pixels. As noted in Sect.~\ref{sec:struct},
the cluster may not even have a well-defined half-light radius.
Secondly, equation~(\ref{eq:virial}) gives the \emph{total} virial mass out to
some large radius, so comparing this dynamical mass with the luminosity
within a radius of, say, 50 pixels may not give a realistic picture of
the extent to which the luminous and dynamical masses agree.  We therefore 
decided to perform a more detailed modeling of the cluster's structure, as
described in the following:

  First, we modeled the cluster density distribution using the projected,
azimuthally-averaged intensity profile in each passband, measured
out to 50 pixels. We interpolated through the intensity bump from the
companion cluster in the northeast using a power law fit to the profile
on either side of it. The corrected profile was then assumed to come
from a spherical cluster with a three-dimensional density profile,
$\rho(r)$, plus some unknown light contamination from foreground and
background stars. In addition to the background subtraction done prior to
photometric analysis, various levels of background light were uniformly
subtracted from the profile to give a range of density solutions,
ranging from the background level measured at the 50th pixel down to
zero background subtraction.  With the largest of these subtracted
backgrounds, the pure-cluster intensity went to zero at the 50th pixel,
as if the cluster had an edge there.

The 3D density profile $\rho(r)$ was then determined from this projected
profile by assuming that the cluster was made from a superposition
of equal-density onion-skin shells, one at each pixel. In fact, the
pixels of the measurements were interpolated linearly onto a finer grid
with four times as many pixels, to get a better accuracy in the final
density. The line-of-sight depth through each of these interpolated
shells is known from the spherical symmetry assumption, so we began at
the outer pixel where the intensity came only from the outer shell and
obtained the density there. This outer density was determined from
the ratio of the intensity to the line-of-sight depth of the outer shell.
The density in the next-inner shell was determined by first subtracting
the intensity at this position coming from the outer shell, using the
appropriate line of sight depth and density there, from the observed
intensity, and then dividing this difference by the line of sight depth
through the next-inner shell. In this way, we could work from the outside
to the inside and determine the 3D density of each shell. The result
is the 3D density profile inside the cluster.

\begin{figure}
\plotone{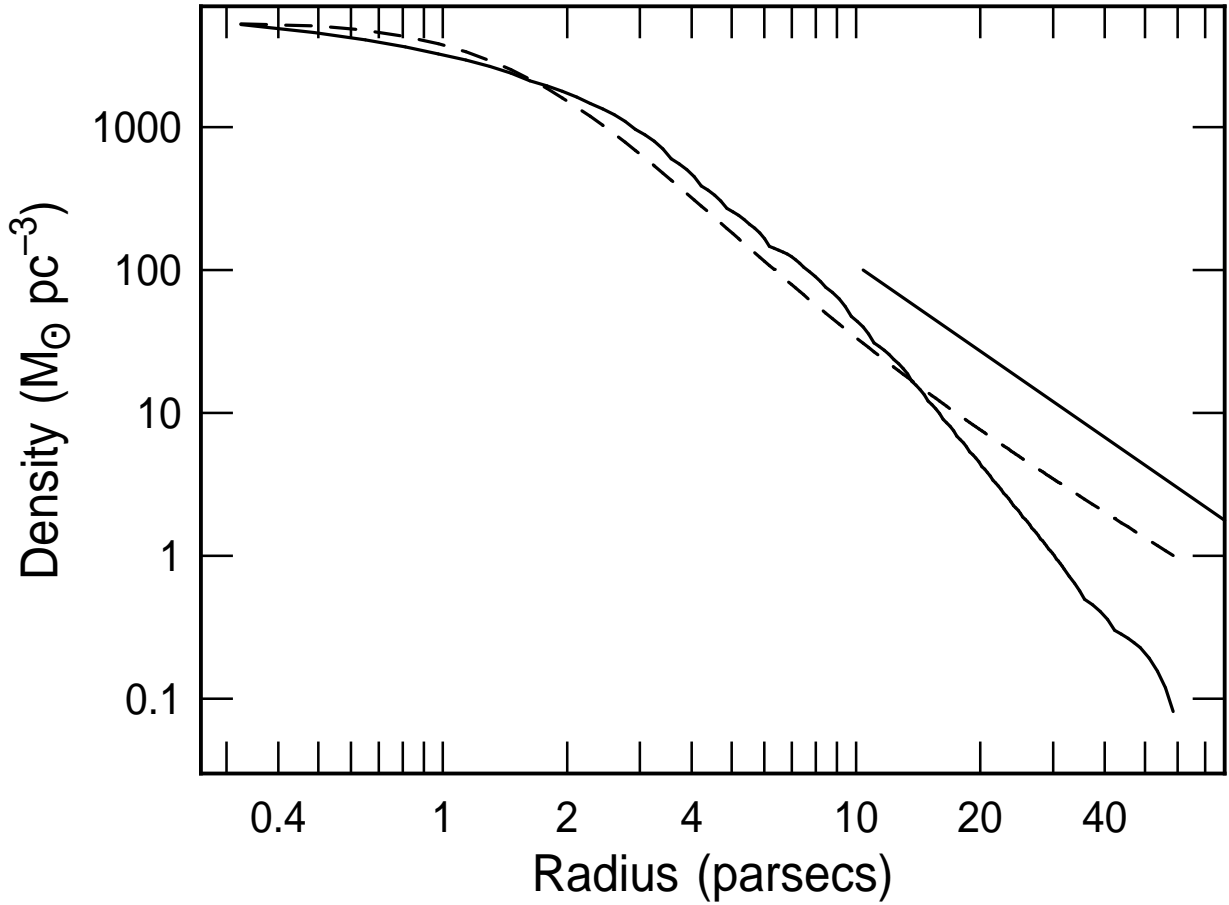}
\figcaption[f9.eps]{\label{fig:3ddens} 
The best-fit solution for the three-dimensional density of the cluster
(solid line) and the corresponding isothermal profile are plotted as
functions of radius.  The straight line has a slope of $-2$. The pixel
scale is 1.3 pc., so the fitted points at small radii are unresolved.
The observed cluster is close to isothermal in the inner regions, but it
becomes cooler in the outer regions where the density drops off faster
than the inverse square of radius.
}
\end{figure}

  Figure \ref{fig:3ddens} shows the fitted 3D density profile
of the cluster determined from the $V$-band projected intensity profile
using a background subtraction that was tuned to give the densities at the
largest few radii a smooth continuation. This will be called the best-fit
solution. Larger subtractions caused the density profile to drop suddenly
at the 50 pixel edge, and smaller subtractions caused it to turn up.
The dashed line in Figure \ref{fig:3ddens} is a solution to the density
structure of an isothermal cluster using the same velocity dispersion
as the average determined from the observations.  The straight dashed
line has a slope of 2, which is the expected slope for an isothermal
cluster at large radius.

The fitted 3D density profile was used in the equation of hydrostatic
equilibrium in order to determine the one-dimensional velocity dispersion,
$v_1(r)$, in each shell.  This dispersion satisfies the equation
\begin{equation}
  d \rho v_1^2/d r = - \rho GM/r^2,
  \label{eq:hydro}
\end{equation}
for mass as a function of radius, $M(r)$, obtained from the density
solution, $M(r)=\int_0^r 4\pi r^2 \rho(r)dr$.  To solve (\ref{eq:hydro}),
we need to know the velocity dispersion at the edge of the cluster
to give the boundary condition on pressure there. We assumed two
cases: zero dispersion at the edge, corresponding to zero pressure,
and a dispersion equal to the average in the cluster, determined from
the fitted $\left(M/R\right)^{1/2}$ at the cluster edge, as given by
the density profile. The run of dispersion, $v_1(r)$ was determined by
integrating from the outside in.  Once this dispersion solution was
obtained, the square of the dispersion was averaged with a weighting
proportional to the shell mass. The square root of this weighted average
then gives the rms dispersion in the whole cluster, as would be observed
with a slit spectrograph that covers it all.  This final dispersion
makes the reasonable assumption that the flux-weighted sum of Gaussian
line profiles from sub-components of a total cluster is approximately
equal to a Gaussian line profile itself, and that the dispersion of this
summed profile is equal to the weighted quadratic sum of the dispersions
of the components.

The absolute calibration for the density and mass now comes from the ratio
of the observed, $v_x$, to the modeled, $v_1$, one-dimensional velocity
dispersions. The absolute mass is $(v_x/v_1)^2\times\left(1.3\;{\rm
pc}\right)/G$ multiplied by the program fitted mass, which is in units
of photon counts.  The absolute density is $(v_x/v_1)^2/\left(1.3\;{\rm
pc}\right)^2/G$ multiplied by the program fitted density.  Here, 1.3 pc
is one pixel.  For the best-fit density solution with an edge dispersion
equal to the average, the mass out to the 50 pixel=65 pc radius becomes
$1.67\times10^6$ M$_\odot$, and the central density is $5.3\times10^3$
M$_\odot$ pc$^{-3}$. With no background subtraction and an average edge
dispersion, the peripheral density is greater and the central density
smaller, $3.8\times10^3$ M$_\odot$ pc$^{-3}$, to give the same observed
velocity dispersion, but the mass inside 50 pixels is about the same.

The case with zero velocity dispersion at the edge is not physical
but it is interesting to compare with the results given by equation
\ref{eq:virial}, which is for an isolated cluster with zero pressure
at the boundary.  Our masses for this zero-dispersion case were
systematically larger than the masses for the average-dispersion
cases, particularly for the models in which there was no background
subtraction. Compared to the best-fit mass above of $1.67\times10^6$
M$_\odot$, the zero edge-dispersion masses were $1.78\times10^6$ M$_\odot$
and $2.81\times10^6$ M$_\odot$ for the best-fit and no-background
subtraction density fits, respectively.  The reason why equation 
\ref{eq:virial} gives a larger mass is that it effectively includes all 
of the mass out to some zero-pressure boundary, even if it is beyond 50 
pixels, but the 3D fit includes only the mass inside 50 pixels. Also,
equation \ref{eq:virial} assumes that the cluster has a well-defined
half-mass radius while our 3D fit derives the mass directly from the 
luminosity profile.

% Thus the mass of
% $\sim3.0\times10^6$ M$_\odot$ found from equation \ref{eq:virial} using
% the observed half-light radius and velocity dispersion should be taken to
% represent some measure of the full cluster mass, out to the far regions
% even beyond our measurement points. This is where the cluster presumably
% blends into the surrounding disk.   The smaller mass that comes from
% the fitted 3D density should be the one compared with the luminous mass
% inside 50 pixels, as determined from population synthesis models below.

  In summary, the $V$-band radial intensity profile was converted to a
3D density profile using a reasonable assumption involving the level of 
background contamination, and this density profile was used to find a 
velocity dispersion profile assuming hydrostatic equilibrium with a 
dispersion at the edge equal to the average obtained from the density fit.  
The weighted average of this dispersion was then compared with the 
observed dispersion to give the absolute calibration for mass and density. 
The result is a cluster mass inside the 65 pc radius equal to 
$\sim1.7\times10^6$ M$_\odot$ and a central cluster density of 
$\sim5.3\times10^3$ M$_\odot$ pc$^{-3}$.  A similar procedure was applied 
to the other passbands with the result that the fitted mass increases 
slightly with wavelength.

The fit to the velocity dispersion profile gives a result that decreases
with radius by a factor of 0.6 and 0.8 from the center to the mid-radial
points for the best-fit and no-background subtraction solutions,
respectively. This decrease is also evident from the difference in Figure
\ref{fig:3ddens} between the fitted 3D density profile and the isothermal
profile, considering that a steeper profile implies a smaller local scale
height and a smaller dispersion for comparable acceleration.  Thus we
predict that the dispersion in the center of the NGC 6946 cluster is
slightly higher than the dispersion at 10 to 20 pc radius. Our solution at
larger radii becomes uncertain because of the unknown starlight background
and the unknown outer boundary condition for the velocity dispersion.

\subsubsection{Luminous mass}

  The dynamical mass estimate of $\sim1.7\times10^6$ M$_\odot$ may be 
compared to a \emph{photometric} estimate, based on the luminosity of 
the cluster and a M/L ratio from population synthesis models. Such a
comparison can potentially provide constraints on the stellar IMF of 
the cluster.  For the following discussion we adopt a cluster age of 
$15\pm5$ Myr, as derived in Sect.~\ref{sec:phot} and in \citet{eel00}.
Using the $V$ magnitude measured through the $r=50$ pixels aperture and
correcting for a galactic foreground extinction of $A_B=1.48$ mag, the
absolute cluster luminosity is $M_V = -13.2$ for a distance modulus of 28.9.
This also includes a correction of $+0.1$ mag to the magnitude listed
in Table~\ref{tab:phot} to account for the fact that an aperture
correction of $-0.1$ mag from $r=0\farcs5$ to $r=\infty$ is implicit in
the Holtzman et al.\ calibration, while our measurement has already been
performed through a large aperture. Thus, $M_V=-13.2$ represents the
luminosity of the cluster \emph{out to $r=50$ pixels (65 pc)}, which is what
should be compared to the dynamical mass derived in Sect.~\ref{sec:mdyn}.

  Using 1996 versions of the Bruzual \& Charlot population synthesis
models \citep[from][]{lei96}, we can estimate the expected luminosity per 
unit mass for a 
given cluster age and stellar initial mass function (IMF). The
Bruzual \& Charlot models are computed for a \citet{salp55} and a 
\citet{scalo86} IMF, the latter approximated as a composite of 6 power-law
segments.  Both extend from 0.1 \msun\ to 125 \msun , but the Scalo IMF 
has a steeper slope than the Salpeter IMF over most of the mass range, 
resulting in a higher mass-to-light ratio.  For an age of $15\pm5$ Myr, 
the models predict $M_V (1 \, \msun) = 2.36\pm0.3$ for the Scalo 
IMF and $M_V (1 \, \msun) = 1.58\pm0.4$ for the Salpeter IMF, 
respectively.  This corresponds to a total mass of 
$1.68\pm0.46\times10^6\msun$ for the Scalo IMF and 
$0.82\pm0.30\times10^6\msun$ for a Salpeter IMF. These numbers
agree fairly well with the dynamical mass within $r=50$ pixels of 
$1.7\pm0.9\times10^6\,\msun$ (Sect.~\ref{sec:mdyn}), although a somewhat 
steeper than Salpeter IMF is preferred.

Alternatively, we may consider an IMF of the form
\begin{equation}
  dN = \left(1-e^{-(M/M_0)^2}\right) M^{-\gamma - 1} dM
  \label{eq:paresce}
\end{equation}
 with $M_0 = 0.4 \, \msun$. 
For $\gamma=1.35$
it approaches a Salpeter function at high masses, but has a shallower
slope for $M < M_0$, making it similar to the IMF reported for old globular 
clusters by \citet{par00}.  If the function is normalized to the 
pure-Salpeter function at high mass to give the same cluster luminosity, 
then it has only 0.67 times as much mass as the Salpeter-only function 
down to $0.1\,\msun$, or $0.55\pm0.20\times10^6$ \msun . This 
is somewhat lower than the dynamical mass estimate, although an IMF of 
the form (\ref{eq:paresce}) is still within the $\sim1\sigma$ error 
margins.

\begin{figure}
\plotone{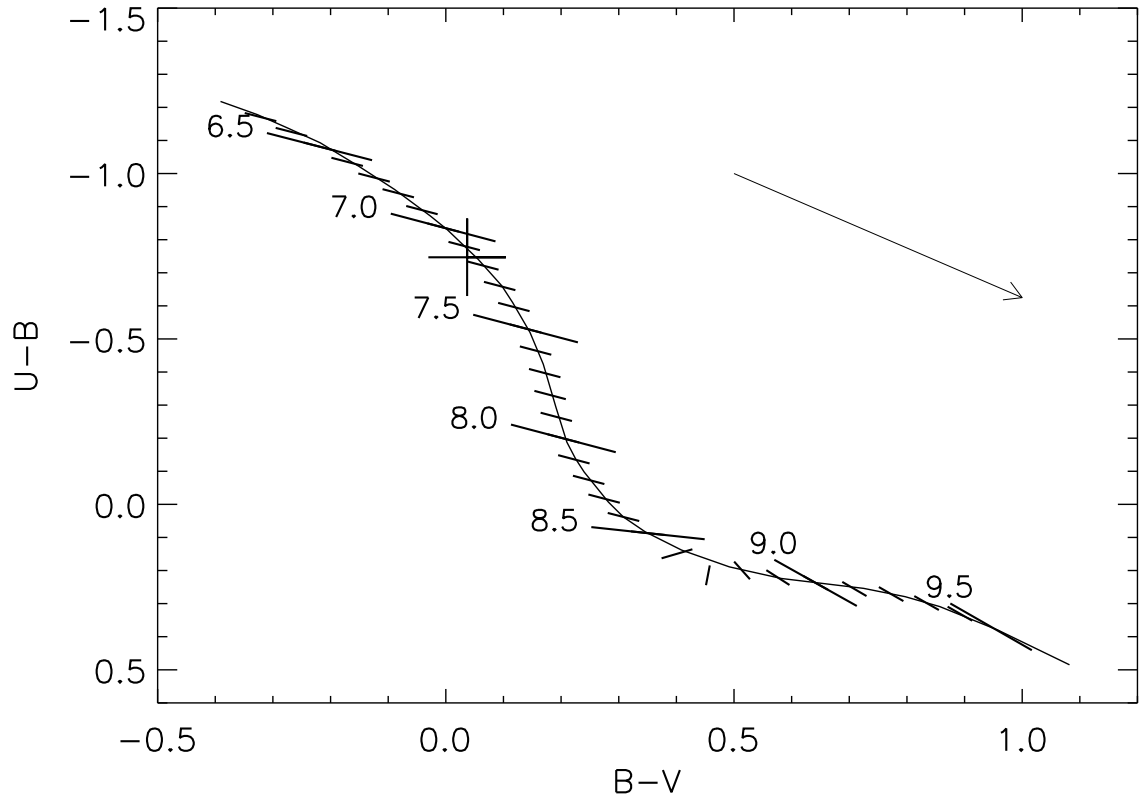}
\figcaption[f10.eps]{\label{fig:bv_ub}(\bv, \ub) two-color diagram showing
the \citet{gir95} S-sequence and the colors of the young globular
cluster in NGC~6946 (large $+$ symbol). Log(age) values are indicated along 
the S-sequence.  The arrow indicates a reddening of E(\bv ) = 0.5.}
\end{figure}

  Although we have corrected for galactic foreground extinction, some 
extinction may still be present within NGC~6946 itself.  \citet{eel00}
suggested that the extinction may vary substantially across the star forming 
region surrounding the young globular.  In Fig.~\ref{fig:bv_ub} we show 
a $\bv, \ub$ two-color diagram with the \citet{gir95} S-sequence 
and a cross indicating the cluster colors corrected for foreground 
extinction. The S-sequence is basically a fit to the colors of LMC 
clusters. The young globular in NGC~6946 lies almost perfectly on the 
S-sequence, but the reddening vector is nearly parallel to the S-sequence 
at this location and the cluster could
be considerably younger if additional absorption is present.
Furthermore, around $10^7$ years, cluster colors do not really change
as smoothly with age as indicated by the S-sequence and ages derived
on the basis of $UBV$ colors should only be taken as approximate
\citep{gir95}.  Assuming an additional $A_B = 0.5$ within NGC~6946, the 
cluster colors would correspond to an age of only $\sim5\times10^6$ yr 
(according to the \citet{gir95} calibration) and the $V$-band luminosity per
unit mass predicted by the Bruzual \& Charlot models would increase 
by $\sim1.1$ mag (for Salpeter IMF).  Although a correction for additional 
reddening would also make the absolute magnitude 0.4 mag brighter in $V$, 
this is not enough to account for the decrease in mass-to-light ratio due 
to lower age.  If the cluster is subject to additional reddening in 
NGC~6946, the net result would therefore be a \emph{decrease} in the 
photometric cluster mass estimate, while the dynamical mass would remain 
unaffected as long as the relative radial profile is the same.

% !!! I added this last phrase about radial profile above.

  An additional uncertainty comes from the distance to NGC~6946.
Although \citet{ksh00} list a mean distance modulus of 28.9 for the
NGC~6946 group and adopt this as the distance of NGC~6946 itself, they
actually obtain a distance modulus of 29.15 for NGC~6946. This would
increase our estimate of the half-mass radius of the cluster to
20 pc and the dynamical mass from equation \ref{eq:virial}
to $3.5\pm1.9\times10^6\msun$. The fitted mass from the 3D density
profile would increase by the same linear factor and become
$2.7\times10^6\msun$. 
% !!! for this I use a distance 1.59 times larger from the mag difference
% then I use the spitzer mass x 1.59 to get 2.6x1.59=4.1.
% My fitted mass also scales with the first power of distance so I would
% increase that from 1.7E6 to 2.7E6. 
Also at this distance, the absolute $V$ magnitude would be $-13.6$
and the resulting photometric mass $2.0\times10^6\msun$ or
$1.0\times10^6\msun$ for Scalo or Salpeter IMFs. 

  In conclusion, our data are compatible with any of the three
stellar IMFs considered here, although a somewhat steeper IMF than
Salpeter is preferred. An IMF with Salpeter slope down to $\sim0.4\,\msun$
and a log-normal shape below this limit is still within the
error limits, but it is clear that our data do \emph{not} favor a
top-heavy IMF with any significant lack of low-mass stars.

% \subsection{Gas near the cluster}

\subsection{Central density}

  The central density $\rho_0$ of the cluster can be estimated from the 
central $V$-band surface brightness $\sigma_0(V)$ in mag arcsec$^{-2}$, 
mass-to-light ratio $M/L$ and core radius in pc \citep{pk75,wb79}:
\begin{equation}
  \rho_0 = 
  \frac{3.44\times10^{10}}{P r_c} 
   10^{-0.4 \sigma_0(V)} (M/L) \, \msun \, \mbox{pc}^{-3}
   \label{eq:dens}
\end{equation}
  where $P \approx 2$. Assuming a light profile of the form 
(\ref{eq:els87}) with $\xi=2.1$ and an extinction-corrected $V$-band 
magnitude of 16.0 within 20 pixels (Table~\ref{tab:phot}), the central 
surface brightness is $\sigma_0(V) = 12.3$ mag arcsec$^{-2}$. Here we
have used the $r=20$ pixels aperture for reference in order to avoid
extrapolation of the model profile to larger radii.  Measuring
the light directly on the image gives $\sigma_0(V) = 13.2$ mag arcsec$^{-2}$
within the central $r=0.5$ pixels.  The direct measurement
is expected to underestimate the central surface brightness because the
central cusp of the profile is smeared by the finite resolution of the
PC. For the same reason, the estimate of 
$\rho_0 = 5.3\times10^3 \, \msun\,{\rm pc}^{-3}$ for the central density from
Sect.~\ref{sec:mdyn} is also likely to be an underestimate.  In the 
following we adopt $\sigma_0(V) = 12.3$ mag arcsec$^{-2}$ as our best 
estimate of the central surface brightness.

% !!! Soeren, you say a mass of 1.5E6 below, and the fit now prefers 1.7E6.
% do you want to reword this below? or do you have another reason to pick 1.5E6? 
  For a mass of $(1.7\pm0.9)\times10^6\,\msun$ and absolute $V$ 
magnitude $M_V=-13.2$, the mass-to-light ratio is $M/L = 0.11\pm0.06$.  
Inserting in equation~(\ref{eq:dens}), the central density is then 
$\rho_0 = (1.7\pm0.9)\times10^4 \,\msun\,{\rm pc}^{-3}$. We can also use
the population synthesis models instead of the dynamical mass to derive
$M/L$ ratios. For an IMF of the form (\ref{eq:paresce}) with $M_0 = 0.4$ 
and $\gamma = 1.35$, we get a $M/L$ ratio of 0.039 and 
$\rho_0 = 6.2\times10^3 \,\msun\,{\rm pc}^{-3}$, which may be considered 
a lower limit.  In any case, the central density is on the order of 
$10^4\,\msun\,{\rm pc}^{-3}$ and thus similar to that of the densest 
stellar clusters in the Milky Way, such as Monoceros R2 and the Trapezium 
cluster \citep{carp97,pros94} and to the R136 cluster at the center of the 
30 Dor complex in the LMC \citep{camp92}, but the total number of stars 
and physical dimensions of n6946-1447 are much larger.

\section{Discussion}

  With a total mass somewhere around $10^6\, \msun$, n6946-1447 is
about an order of magnitude more massive than the most massive
young cluster in the LMC, NGC~1866 \citep{fis92} and many orders of
magnitude more massive than typical open clusters in the Milky Way.
It is, however, comparable to the most massive clusters in merger and 
starburst galaxies such as the Antennae \citep{zha99}. This clearly
illustrates that, although such clusters are \emph{mostly} observed
in merger galaxies and other starburst environments, they can also form 
far from the nucleus in quite normal, apparently undisturbed disk 
galaxies. Violent interactions such as galaxy collisions may help to 
create an environment that is favorable for formation of massive clusters, 
but such events are evidently not a \emph{necessary} condition.  The 
density near the cluster center seems to be similar to dense star forming 
regions in the Milky Way. Thus, the basic star forming mechanism at
work may well be the same, although proceeding at a much larger scale
in n6946-1447.

  It is also of interest to compare n6946-1447 with two of the most luminous 
old globular clusters in the Milky Way, $\omega$ Cen and 47 Tuc.  From 
their dynamical properties, \citet{mey86} estimate total masses for the 
two clusters of $2.9\times10^6\,\msun$ and $1.3\times10^6\,\msun$, 
respectively. Considering that a significant fraction of the mass may be
located beyond 65 pc, this makes n6946-1447 comparable in mass to these 
two old globular clusters.  47 Tuc has a relatively compact core with 
$r_c = 0.46$ pc, while $\omega$ Cen is a quite loosely structured cluster 
with $r_c = 3.8$ pc \citep{har96}. The effective radii of the two clusters are 
3.5 pc and 6.2 pc. Our estimate of the core radius for n6946-1447 of 
1.3 pc is intermediate between 47 Tuc and $\omega$ Cen, while the 
effective radius is larger than for either of the two old globulars.
As argued in Sect.~\ref{sec:struct}, the outer parts of the cluster may
eventually be stripped away and this might decrease the effective radius
of n6946-1447 over time.  
% !!! added "over time" above
The rotation curve of NGC~6946 indicates a mass of
$\sim3\times10^{10}\,\msun$ within the location of n6946-1447 at about
4.5 kpc from the center \citep{car90}. For a cluster mass of 
$1\times10^6\,\msun$, this corresponds to a tidal radius of about 70--100 pc
\citep{king62,keen81}. This number depends only weakly on the exact masses of
the galaxy and cluster, but assumes a homogeneous gravitational field.
In practice, passages near giant molecular clouds, through spiral arms
etc.\ may further contribute to stripping of stars from the cluster.

  One outstanding question has been whether or not young massive
clusters will be able to survive for any considerable amount of time.
Here we find that the dynamical mass estimate of n6946-1447 agrees well
with to the mass predicted by various population synthesis models.
Formally, a slightly steeper than Salpeter IMF is preferred, but a Salpeter
IMF with a lower mass limit of $0.1\,\msun$ is within the error margins.
The cluster IMF could even have a Salpeter slope down to 0.4\msun\ and a
log-normal behavior below this limit, but our data do not support a 
``top-heavy'' IMF with a significant lack of low-mass stars. A Salpeter IMF
with a lower mass limit of $2\,\msun$, for example, would give a total
cluster mass of only $\sim2\times10^5\,\msun$. 
% Similar 
% conclusions have been reached for young massive clusters in NGC~1569 and 
% NGC~1705 \citep{hf96a,hf96b,stern98} and for the luminous cluster `F' in 
% M82 \citep{sg00}.  

\section{Summary and conclusions}

  We have presented new HST / WFPC2 and Keck / HIRES data for an extremely
luminous young star cluster in the nearby spiral galaxy NGC~6946.  
Within an $r=50$ pixels (65 pc) aperture, the integrated cluster luminosity 
is $M_V=-13.2$, making the cluster luminosity similar to that of young 
`super-star clusters' observed in starburst galaxies. It is certainly the 
most luminous star cluster known in the disk of any normal spiral 
galaxy.  From the PC images we find that the cluster has a compact core 
with a core 
radius of about 1.3 pc, but is surrounded by an extended envelope. At large 
radii, the luminosity profile as a function of radius is well modeled by 
a power-law function with an exponent close to $-2$, similar to the profile
of young clusters in the LMC \citep{els87} and to a \citet{king62} profile
with a large tidal radius. We estimate the half-light radius $\reff$ to be
about 13 pc, but this estimate is uncertain because of the extended halo 
that surrounds the cluster. However, it agrees well with a previous 
estimate based on ground-based data of $\reff \sim 11$ pc \citep{eel00}.

% !!! slight rewording below. 

  From the Keck / HIRES high-dispersion spectra we estimate a velocity
dispersion of $10.0\pm2.7$ km/s for the cluster.  From a detailed modeling 
of the density profile of the cluster, we find that this implies a dynamical 
mass within 65 pc of $1.7\pm0.9\times10^6\,\msun$. 
Bruzual \& Charlot population synthesis models predict a mass of
about $1.7\times10^6 \,\msun$ for a \citet{scalo86} stellar IMF and 
$0.8\times10^6 \,\msun$ for a \citet{salp55} IMF from 0.1 -- 125 \msun . 
If the IMF is Salpeter down to $0.4\,\msun$ and log-normal below this
mass, as seen in old globular clusters \citep{par00} then the predicted
cluster mass is $0.55\times10^6\,\msun$.  Comparing the photometric and 
dynamical mass estimates and taking the associated uncertainties into
account, we find that the IMF presumably contains at least as much mass 
in low-mass stars as a Salpeter law (but a turn-over at $0.4\,\msun$ is
within the uncertainty limits) and the cluster will most likely remain bound, 
although it may lose much of its outer envelope to tidal forces.  The cluster 
is comparable in mass to the most massive clusters in the Antennae galaxy 
and an order of magnitude more massive than the young LMC cluster NGC~1866.  
The central density of the cluster is about $10^4\,\msun$ pc$^{-3}$,
comparable to the densest star forming regions in the Milky Way such as 
the Trapezium cluster in Orion.  
  
\acknowledgments

  Support for Program number GO-8715 was provided by NASA through
grants GO-08715.02-A and GO-08715.05-A from the Space Telescope Science 
Institute, which is operated by the Association of Universities for 
Research in Astronomy, Incorporated, under NASA contract NAS5-26555. 
Some of the data were obtained at the W.M. Keck Observatory, which is 
operated as a scientific partnership among the California Institute
of Technology, the University of California and the National Aeronautics 
and Space Administration.  The Observatory was made possible by the 
generous financial support of the W.M. Keck Foundation.
JB and SSL acknowledge support by National Science Foundation grant 
number AST9900732. Yu.E. is grateful for support by Russian  FBR  
grants  00-02-17804 and 00-15-96627. We thank E.\ Alfaro and P.\
Battinelli for their help and the referee, Dr.\ Brad Whitmore, for
a number of useful comments and suggestions.

\clearpage

% \epsfxsize=10cm
% \epsfbox{fov.ps}

% \epsfxsize=8cm
% \epsfbox[127 255 505 548]{spec_opt.ps}
% \epsfxsize=8cm
% \epsfbox[127 255 505 548]{spec_ir.ps}

% \epsfbox{ygc.eps}

% \epsfxsize=8cm
% \epsfbox{rprof.ps}
% \epsfxsize=8cm
% \epsfbox{rprof2.ps}

% \noindent\epsfbox{ccor.eps}
%

% \epsfxsize=10cm
% \epsfbox{demo.ps}

% \epsfxsize=12cm
% \epsfbox{vdisp.ps}

% \epsfxsize=10cm
% \epsfbox[84 121 505 435]{dens3d.ps}

% \epsfxsize=10cm
% \epsfbox{bv_ub.ps}

\clearpage

\begin{table}
\begin{center}
\caption{\label{tab:tmpl}Template stars used for cross-correlation.
}
\begin{tabular}{ll}
\tableline\tableline
Star     & Type \\ \tableline
HR 37   & K5III \\
HR 97   & G5III \\
HR 207  & G0Ib \\
HR 213  & G8II \\
HR 690  & F7Ib \\
HR 861  & K3Ib \\
HR 1009 & M0II \\
HR 8412 & G5Ia \\
HR 8692 & G4Ib \\
HR 8726 & K5Ib \\
HR 9053 & G8Ib \\
\tableline
\end{tabular}
\end{center}
\end{table}

\begin{table}
\begin{center}
\caption{\label{tab:phot}Comparison of PC and ground-based
photometry for the young globular cluster in NGC~6946. No correction for
Galactic extinction has been applied. Note that the magnitudes listed
here include an implicit $-0.1$ mag aperture correction, which is
only valid for a point source observed through a $0\farcs5$ (11 pixels)
aperture.
}
\begin{tabular}{lcccc}
\tableline\tableline
Aperture  & $V$   & \ub      & \bv   & \vi   \\ \tableline
5 pix  & $17.944\pm0.002$ & $-0.443\pm0.005$ & $0.488\pm0.004$ & $1.217\pm0.002$ \\
10 pix & $17.458\pm0.002$ & $-0.447\pm0.004$ & $0.462\pm0.004$ & $1.166\pm0.002$ \\
20 pix & $17.042\pm0.002$ & $-0.461\pm0.004$ & $0.439\pm0.004$ & $1.134\pm0.002$ \\
30 pix & $16.880\pm0.002$ & $-0.475\pm0.005$ & $0.420\pm0.004$ & $1.122\pm0.002$ \\
50 pix & $16.683\pm0.002$ & $-0.491\pm0.006$ & $0.410\pm0.004$ & $1.103\pm0.003$ \\
Ground  & 16.91 & $-0.47$  & 0.37  & 1.10 \\
\tableline
\end{tabular}
\end{center}
\end{table}

\begin{table}
\begin{center}
\caption{\label{tab:vd}Estimates of the velocity dispersion of the
young globular for different echelle orders and template star luminosity
classes. The errors are computed as the standard deviation of all
velocity dispersion estimates for a given echelle order and luminosity
class.  Prefixes 'OPT' and 'IR' denote the optical and near-infrared
spectrograph setups.
}
\begin{tabular}{lrrrr}
\tableline\tableline
Order    & Pixels    & \multicolumn{3}{c}{Velocity disp. [km/s]} \\ 
         &           &   I (7 stars)  & II (2 stars) & III (2 stars) \\ \tableline
OPT--29 &  50--2000 & 12.8$\pm$2.8 &  8.9$\pm$2.3 & 12.4$\pm$2.2 \\
OPT--34 &  50--2000 & 11.8$\pm$1.6 & 13.0$\pm$2.4 & 16.2$\pm$1.6 \\
OPT--37 &  50--2000 & 10.4$\pm$0.9 &  9.1$\pm$1.3 & 11.0$\pm$0.6 \\
IR--3   &  50--2000 & 15.1$\pm$1.4 & 14.4$\pm$1.5 & 15.5$\pm$0.2 \\
IR--4   & 500--2000 &  8.7$\pm$1.6 &  7.3$\pm$1.0 &  8.8$\pm$0.7 \\
IR--7   &  50--2000 &  5.6$\pm$0.7 &  6.5$\pm$0.8 &  8.1$\pm$0.6 \\
IR--10  &  50--2000 &  9.9$\pm$0.8 &  8.8$\pm$1.1 & 10.9$\pm$0.3 \\
IR--11  & 100--2000 &  8.9$\pm$0.8 &  9.2$\pm$0.6 & 10.0$\pm$0.2 \\
IR--14  &  50--2000 & 10.1$\pm$1.7 & 11.2$\pm$1.4 & 12.4$\pm$0.7 \\
\tableline
\end{tabular}
\end{center}
\end{table}

\end{document}